\documentclass[%
 reprint,
 amsmath,amssymb,
 aps,
 prl
]{revtex4-1}
\usepackage{graphicx}
\usepackage{dcolumn}
\usepackage{bm}
\usepackage{color}

\newcommand*{\rtten}[1]{\bm{#1}}
\newcommand*{\rtvec}[1]{\mathbf{#1}}

\begin{document}

\preprint{APS/123-QED}

\title{Stability of Electrodeposition at Solid-Solid Interfaces and Implications for Metal Anodes}

\author{Zeeshan Ahmad}
\affiliation{%
 Department of Mechanical Engineering, Carnegie Mellon University, Pittsburgh, Pennsylvania 15213, USA
}
\author{Venkatasubramanian Viswanathan}
 \email{venkvis@cmu.edu}
\affiliation{%
 Department of Mechanical Engineering, Carnegie Mellon University, Pittsburgh, Pennsylvania 15213, USA
}
 \affiliation{%
 Department of Physics, Carnegie Mellon University, Pittsburgh, Pennsylvania 15213, USA}

\date{\today}

\begin{abstract}
We generalize the conditions for stable electrodeposition at isotropic solid-solid interfaces using a kinetic model which incorporates the effects of stresses and surface tension at the interface. We develop a stability diagram that shows two regimes of stability: previously known pressure-driven mechanism and a new density-driven stability mechanism that is governed by the relative density of metal in the two phases. We show that inorganic solids and solid polymers generally do not lead to stable electrodeposition, and provide design guidelines for achieving stable electrodeposition.
\end{abstract}

\maketitle

Electrodeposition, a process of great practical importance in thin films and metallurgy, has served as a platform for understanding nonequilibrium growth processes and studying morphological instabilities \cite{gamburg2011theory, ross1994electrodeposited}.  Theoretical and experimental investigations have focused on developing a comprehensive understanding of the origin of morphological instability \cite{Barton1962, Chazalviel90, Shraiman84, Voss85} and a rich variety of morphologies including fractal structures have been observed through control of the electrode potential and metal ion concentration \cite{PhysRevLett.56.1260, PhysRevLett.76.4062, PhysRevLett.76.4062, PhysRevLett.61.2558, PhysRevLett.86.256, MatsushitaFractal1984, erlebacher-93-comp}. The study of dendritic growth during electrodeposition has gained renewed interest in light of their importance in the safety issues associated with dendritic short in current Li-ion batteries \cite{aurbach2002short}.  Further, controlling the growth of dendrites during electrodeposition could enable the use of metal anodes especially based on Li which could lead to significantly higher energy density batteries \cite{XuLi2014, oleg2015TMR}. 

Of many possible approaches to control the growth of dendrites, suppression through the use of a solid electrolyte has emerged as the most promising route \cite{Schaefer2012, TikekarNatEnergy2016}. When the liquid electrolyte in contact with metal electrode is replaced by a solid phase, creating a solid-solid system, the interface properties alter the local kinetics of electrodeposition \cite{Monroe2004Effect}. Monroe and Newman analyzed the interfacial stability of Li/solid  polymer electrolyte system within linear elasticity theory and showed using a kinetic model that solid polymer electrolytes with a sufficient modulus are capable of suppressing dendrite growth \cite{Monroe2005Impact}.   However, the propagation of the interface is often accompanied by a change in density of the metal and thus, density is an important order parameter that should affect the stability of electrodeposition at the interface. In the theory of roughening of solid-solid interfaces studied in geological systems, it has been shown that interfacial stability or roughening condition depends the density change at the interface \cite{Angheluta2009}. Furthermore, the stability is determined by a subtle interplay between the density, modulus and the Poisson's ratio. 

In this work, we derive general stability criteria for electrodeposition at solid-solid interfaces using linear stability analysis assuming that the solids are linearly elastic isotropic materials. Based on the stability criteria, we show that there is a new stabilizing mechanism that is determined by density change between the two solids.  Our analysis shows that it is possible to use a soft solid electrolyte provided the partial molar density of the metal is greater in the solid electrolyte as compared to the metal anode.  This mechanism opens up new ways to suppress dendrite growth at Li electrode/solid electrolyte interfaces.  We construct a general stability plot with two parameters, shear modulus ratio and molar volume ratio, and show that two distinct regions of stable electrodeposition are possible.  We find that typical inorganic solid electrolytes have higher shear modulus, but lower molar volume than that required for stable electrodeposition, leading to unstable electrodeposition.  On the other hand, solid polymer electrolytes have higher molar volume but lower shear modulus than the requirement, leading once again to unstable electrodeposition.  Our analysis suggests that a solid electrolyte with a combination of high (low) Li molar volume and high (low) shear modulus  is required for stable electrodeposition.

\begin{figure}[htbp]
{\includegraphics[scale=1.0]{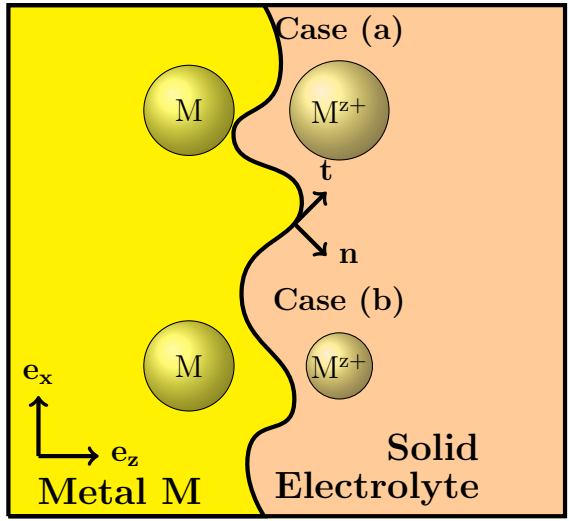}}
\caption{\label{fig:schematic} Schematic of the electrodeposition problem with metal electrode-solid electrolyte interface. The metal surface $z=f(x,t)$ grows on deposition of metal ions, the rate of which is proportional to the current. The local geometry alters the kinetics of deposition at the interface.}
\end{figure}

We study the system of a metal electrode in contact with a solid containing mobile metal ions (solid electrolyte), as shown in Fig. \ref{fig:schematic}. 
This situation is common in electroplating and during charging at metal anodes in batteries. In this process, $\mathrm{M^{z+}}$ ions from the electrolyte are reduced and deposited at the metal electrode as metal atoms according to the reaction:
\begin{eqnarray}\label{eq:reaction}
\mathrm{{M}^{z+}+ze^{-}}\rightleftharpoons\mathrm{M}.
\end{eqnarray}

Based on the operating conditions, this process could lead to stable electrodeposition or morphological instabilities due to uneven deposition of metal ions at the electrode surface. To understand the non-equilibrium growth process and its stability, we need to determine the rate of deposition at the interface. We are interested in the \emph{initiation} of small perturbations at the interface and we will ignore grain boundaries in the solid electrolyte through which these small perturbations may \emph{propagate} after initiation \footnote{The initiation regime has been shown to be most critical for dendrite suppression since dendrites cannot be suppressed if they reach propagation regime. See Ref. \onlinecite{Monroe2004Effect} and references cited in its introduction.}. Experimental studies have also indicated that solid electrolytes need to be prepared without grain boundaries or interconnected pores using dense electrolyte preparation methods like pressure-assisted sintering in order to function in a battery \cite{Ren2015direct}. Nevertheless, we later provide means by which the effect of defects like grain boundaries may be included in the model. The evolution of the metal surface $z=f(x,t)$ can be related to the current density at the interface:
\begin{eqnarray}\label{eq:surfgrowth}
\frac{\partial f(x,t)}{\partial t}\rtvec{e_z}\cdot \rtvec{e_n}=-\frac{iV_{\text{M}}}{\mathrm{z} F}.
\end{eqnarray}
where $\rtvec{e_n}$ is the unit normal pointing from the metal towards the solid electrolyte, $V_{\text{M}}$ is the molar volume of the metal, $F$ is the Faraday constant and $i$ is the current density normal to the interface. The current density, $i$, can be related to the surface overpotential $\eta_s$ through the Butler-Volmer relationship:
\begin{eqnarray}\label{eq:i}
\frac{i}{i_0}= \left[ \exp \left(\frac{\alpha_a \mathrm{z}F \eta_s}{RT} \right)  - \exp \left(-\frac{\alpha_c \mathrm{z}F \eta_s}{RT} \right)  \right].
\end{eqnarray}
Here  $\alpha_a$ and $\alpha_c$ are the charge transfer coefficients associated with anodic and cathodic reactions and $i_0$ is the exchange current density. The Butler-Volmer relationship is known to describe electrodeposition processes well for small surface overpotentials and moderate currents \cite{newman2012electrochemical}.

In our analysis, we consider a constant metal ion concentration at the interface which is a good approximation for solid electrolytes. A large deviation from the average concentration of metal ion will cause local violation of electroneutrality since the anions are generally fixed, resulting in a large energy penalty  \cite{bazant2005, newman2012electrochemical}. Under this assumption, a constant driving force at the interface will result in a uniform surface development without irregularities. However, the local interface geometry affects the driving force for electrodeposition and thereby, the kinetics of metal deposition. Hence it is essential to describe a kinetic relationship that takes into account the local interface geometry. Locally, the electrochemical potential changes due to surface tension and interfacial stresses in a solid. Earlier models used surface tension as the primary stabilizing mechanism against morphological instability. These include the notable works of Mullins and Sekerka on solidification \cite{MullinsSekerka63, MullinsSekerka64} and Barton and Bockris on electrochemical systems \cite{Barton1962}. However, the interfacial stress can have a major influence on the growth morphology in solids \cite{Carter93}. More recently, the effect of mechanical stresses have been incorporated into electrochemical problems \cite{Monroe2004Effect,Tikekar2016}. Here, we will follow the Monroe-Newman approach as it explicitly includes the Butler-Volmer kinetic relationship at the interface. The new kinetic relationship at a deformed interface within this model can be written as:
\begin{eqnarray}\label{eq:inew}
\frac{i_{\mathrm{deformed}}}{i_{\mathrm{undeformed}}}=\exp\left[\frac{(1-\alpha_a)\Delta \mu_{e^-}}{RT} \right] \end{eqnarray}
where $i_{\mathrm{undeformed}}$ is the current density at an undeformed interface given by Eq. (\ref{eq:i}) and $\Delta \mu_{e^-}$ is the change in electrochemical potential of the electron at a deformed interface. It depends on the surface tension and interfacial stresses as \cite{Monroe2004Effect}:
\begin{eqnarray}\label{eq:mu}
\begin{split}
\Delta \mu_{e^-}=&-\frac{1}{2\mathrm{z}}\left(V_\mathrm{{M}} + V_\mathrm{{M^{z+}}}\right) \left(-\gamma \kappa \right.  \\
&\left. + \rtvec{e_n}\cdot [(\rtten{\tau_d^e} - \rtten{\tau_d^s}) \rtvec{e_n}]\right)\\
& + \frac{1}{2\mathrm{z}} \left(V_\mathrm{{M}} - V_\mathrm{{M^{z+}}} \right) \left(\Delta p^e + \Delta p^s \right).
\end{split}
\end{eqnarray}

Here, $V_{\mathrm{M^{z+}}}$ is the molar volume of $\mathrm{M^{z+}}$ in the solid electrolyte, $\gamma$ is the surface tension at the interface, $\kappa$ is the mean curvature at the interface, $\rtten{\tau_d^e}$ and $\rtten{\tau_d^s} $ are the deviatoric stresses at the electrode and electrolyte sides of the interface,  and $\Delta p^e$ and $\Delta p^s$ are the gage pressures at the electrode and electrolyte sides of the interface. Eq. (\ref{eq:mu}) is obtained by calculating the electrochemical potential change $d\mu=(\partial \mu/\partial p) dp$ and using the equilibrium of Eq. (\ref{eq:reaction}). Given the geometry of the interface and material response to resulting strains, it is possible to calculate the local kinetic term and obtain the instantaneous surface growth rate from Eq. (\ref{eq:surfgrowth}). A convenient and sufficiently general choice of the initial geometry to study morphological stability is a sinusoidal perturbation of the interface since the equations of motion can be solved analytically in this case \cite{Monroe2005Impact} and any electrode surface geometry can be expanded as a Fourier series. Consistent with a linear stability analysis, the interface at $z=0$ is perturbed with a perpendicular displacement (i.e. along $\mathbf{e_z}$) of the form $u_z(x,z=0)=\text{Re}\{Ae^{ikx}\}$ with $A\ll 1$. Unlike the Asaro-Tiller formalism \cite{Asaro1972, carter-asaro-1998}, the electrochemical potential change due to strain energy density is of second order and can be neglected in our linear stability analysis. The displacements are assumed to vanish far from the interface i.e. $\lim_{z\to\pm\infty}\rtvec{u}(x,z)=0.$ The traction boundary condition is a tangential force balance at the interface:
\begin{eqnarray}
\rtvec{e_t} \cdot [(\rtten{\tau}_d^{e} - \rtten{\tau}_d^{s})\rtvec{e_n}]=0.
\end{eqnarray}
Using these boundary conditions, bulk force balance: $\text{div } \rtten{\sigma}=0$, and constitutive laws for a linearly elastic isotropic material with shear modulus $G$ and Poisson's ratio $\nu$, $\Delta \mu_{e^-}$ can be computed for every point on the interface. When the values of stresses and surface tension are plugged into the Eq. (\ref{eq:mu}), we obtain  $\Delta \mu_{e^-} = \chi\text{Re}\{Ae^{ikx}\}$ with $\chi=\chi(G_e,G_s,\nu_e,\nu_s,\gamma,k,\mathrm{z},V_{\mathrm{M}},V_{\mathrm{M^{z+}}})$ \footnote{See Supplemental Material for exact expression for $\chi$ and critical shear modulus ratio, and calculations of partial molar volume ratio.} . Stable electrodeposition will occur when current density is out of phase with the perturbation. Equivalently, $\Delta \mu_{e^-}$ should be out of phase with the perturbation (since $1-\alpha_a>0$ in Eq. (\ref{eq:inew})) i.e. $\chi<0$, in which case the the deposition will be faster at the valleys ($A\cos(kx)<0$) than the peaks ($A\cos(kx)>0$), resulting in an even surface growth. Since the sign of $\chi$ determines the stability of electrodeposition, hereafter, we refer to $\chi$ as the stability parameter. This result is similar to that for stability of a material surface against interface migration encountered in fabrication of epitaxial thin films \cite{freund2004thin}.

\begin{figure}[!htbp]
\includegraphics[width=0.45\textwidth]{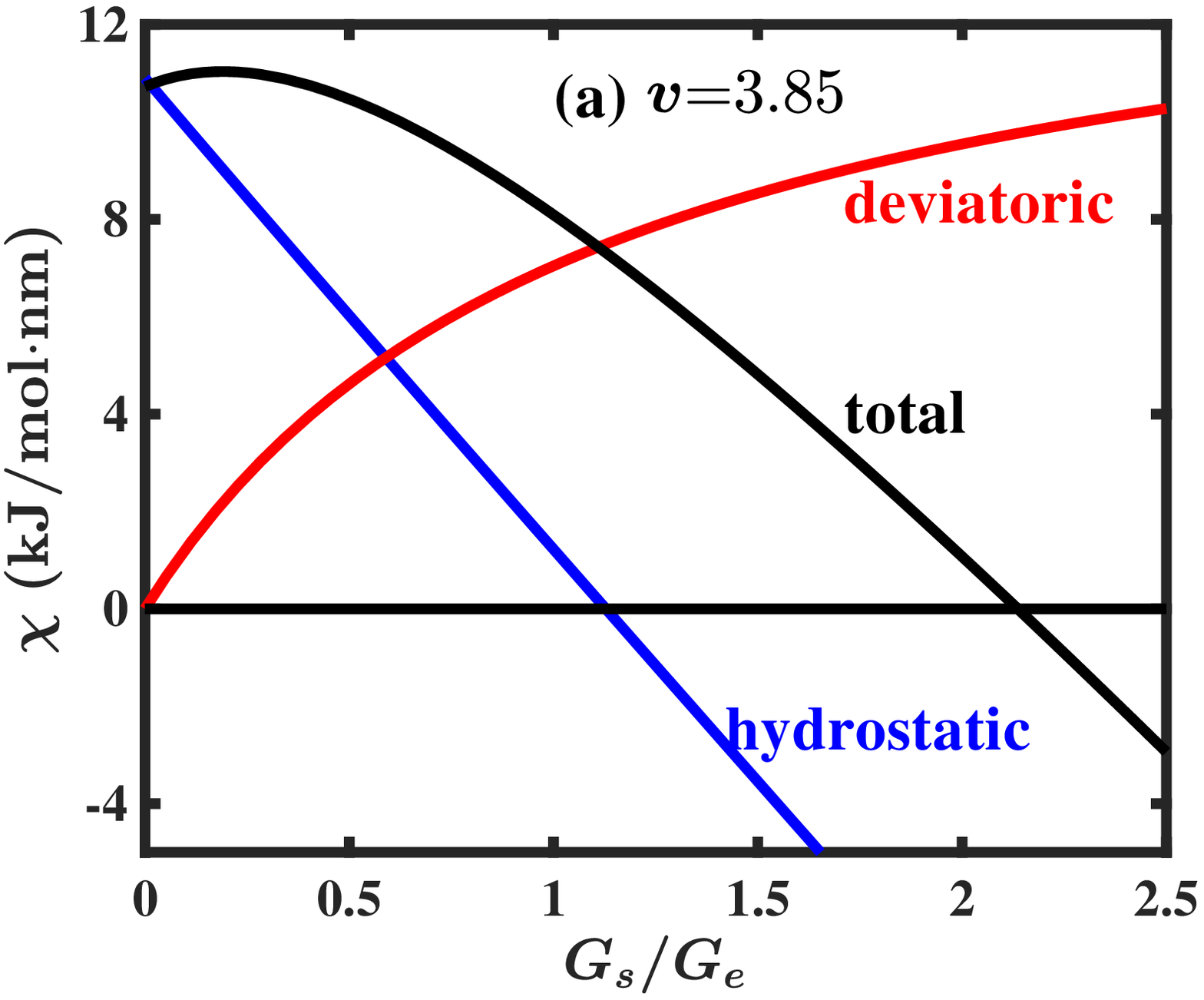}
\includegraphics[width=0.45\textwidth]{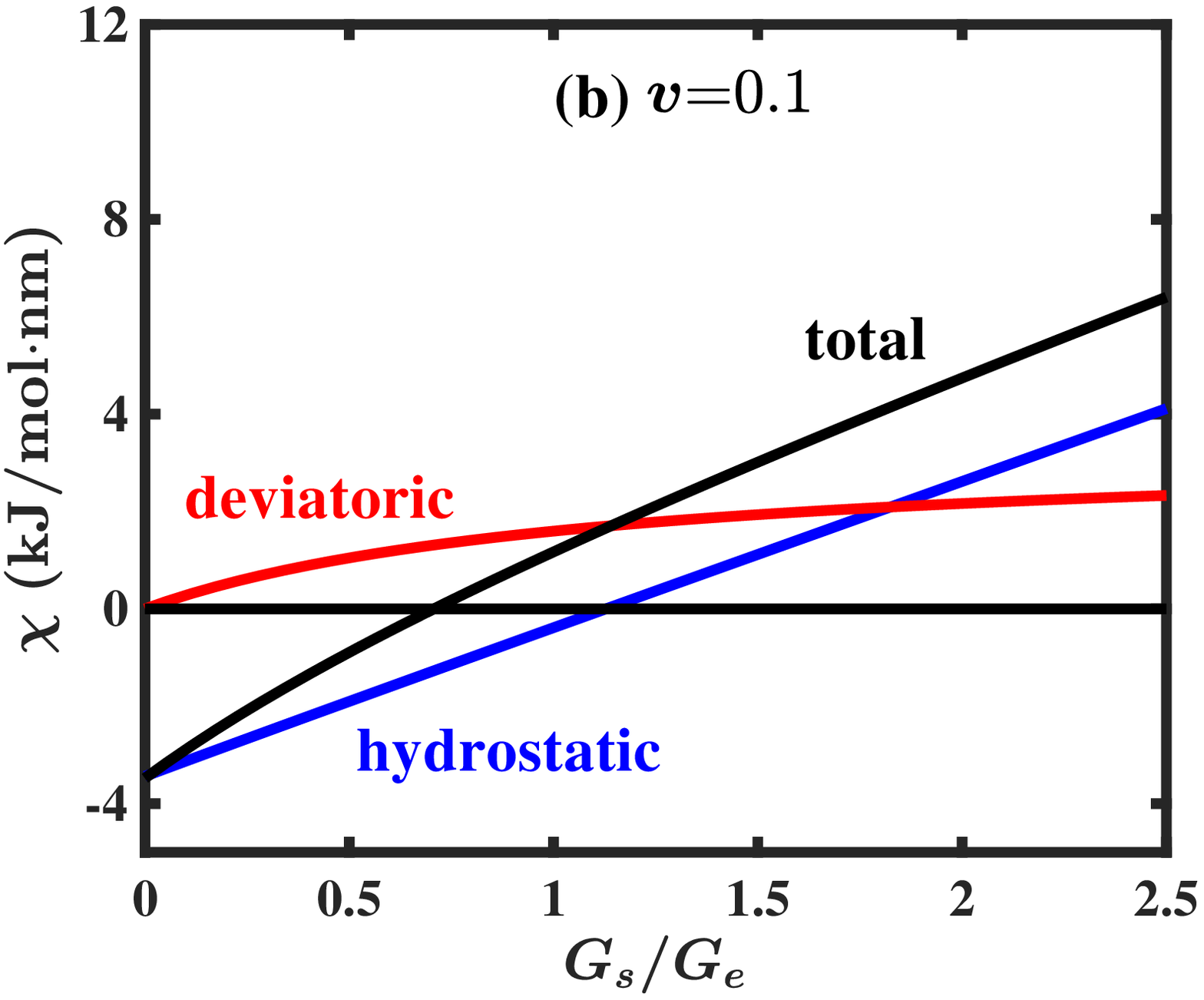}
\caption{\label{fig:mucontri} Contributions of different terms to the stability parameter $\chi$ for (a) $v>1$ and (b) $v<1$ with Li metal electrode. The property values used for the plots were $G_e=3.4$ GPa, $\nu_e=0.42$, $\nu_s=0.33$, $V_{M}=1.3\times 10^{-5}$ $\mathrm{m^3}$/mol, $V_{\mathrm{M^{z+}}}=0.3\times 1.674\times 10^{-4}$ $\mathrm{m^3}$/mol, $k=10^8$/m, $A=0.4/k$ (Ref. \onlinecite{Monroe2005Impact}) for (a) and $v=0.1$ for (b). The surface tension term is evaluated by choosing $\gamma$ as the average surface energy of Li for (100), (110) and (111) planes, giving a value 0.556 J/$\mathrm{m^2}$ \cite{Vitos98surface}. This term is generally small at the wave numbers of perturbation and its contribution has not been shown. The deviatoric stress term is always destabilizing. For $v>1$ the hydrostatic stress term is destabilizing at low $G_s/G_e$ and stabilizing at high $G_s/G_e$ whereas for $v<1$, it is stabilizing at low $G_s/G_e$ and destabilizing at high $G_s/G_e$.}
\end{figure}

Eq. (\ref{eq:mu}) shows that $\Delta \mu_{e^-}$ and hence, $\chi$ consists of contributions from surface tension, hydrostatic and deviatoric stresses. The stabilizing or destabilizing nature of the hydrostatic term depends on the sign of $V_\mathrm{{M^{z+}}}-V_\mathrm{M}$. Therefore, the volume ratio $v=V_\mathrm{{M^{z+}}}/V_\mathrm{M}$ is an important order parameter of the electrodeposition problem. A hydrostatically stressed interface will inhibit growth of dendrites when $v>1$ such as in polymers and viscoelastic liquids with high elastic response, and considerable ion-solvent interactions \cite{marcus2004}. On the other hand, the hydrostatic stress term will be destabilizing for $v<1$ and this is generally the case for inorganic solid electrolytes as we will show later. Fig. \ref{fig:mucontri} shows hydrostatic and deviatoric contributions to $\chi$ for (a) $v>1$ and (b) $v<1$ as a function of the ratio $G_s/G_e$ with Li metal as the electrode. In (a), the hydrostatic contribution is initially positive (destabilizing) and monotonically decreasing with $G_s/G_e$ which results in stability when $G_s/G_e\gtrsim 2.2$ when this term starts to dominate the stability parameter. The scenario reverses for (b) where the hydrostatic stress term is initially negative (stabilizing) and monotonically increasing resulting in stability for $G_s/G_e\lesssim 0.7$. It is worth noting that the deviatoric stress term is always destabilizing. The surface tension term is very small ($<$0.2 kJ/mol$\cdot$nm) at the wave numbers of perturbation of interest and has been ignored in further analysis. However, techniques like  nanostructuring the interface \cite{Wang17} might enhance its contribution to the stability parameter.

\begin{figure}[!htbp]
\includegraphics[width=0.45\textwidth]{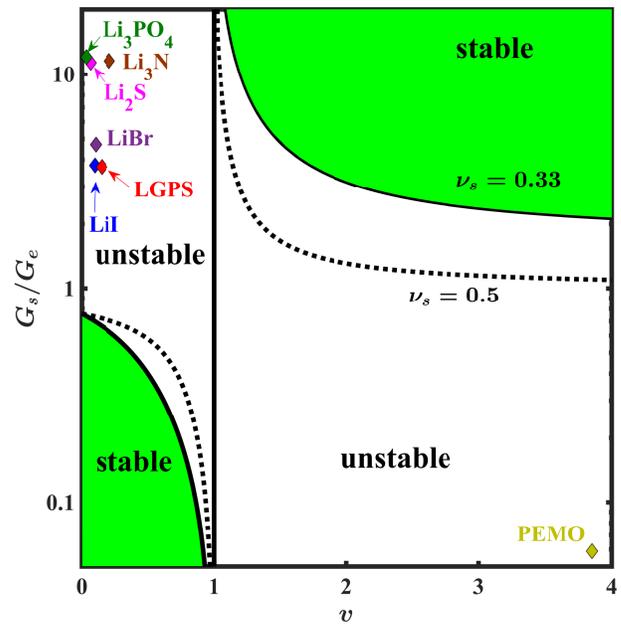}
\caption{\label{fig:phase-diag} Stability diagram showing the range of shear moduli over which electrodeposition is stable and its dependence on the volume ratio $v$ of the cation and metal atom. Regions with stable electrodeposition are shaded green. The critical curves separating stable and unstable regions are plotted using $\nu_e=0.42$ (Li metal) and $\nu_s=0.33$, $0.5$ (incompressible). Several Li solid electrolytes are also plotted in the diagram where the ratio $G_s/G_e$ has been calculated using $G_e=G_{\text{Li}}=3.4$ GPa. For LGPS ($\mathrm{Li_{10}GeP_2S_{12}}$), $V_{\mathrm{Li^+}}$ was calculated from the coordination number whereas for all others, the procedure mentioned in the Supplemental Material \cite{Note2} was used. The solid polymer electrolyte shown is a 10\% by weight solution of PEMO and LiTFSI in glyme \cite{Hafezi2002}.}
\end{figure}

The results from Fig. \ref{fig:mucontri} show that for $v>1$, there exists a critical shear modulus ratio beyond which the electrodeposition is stable ($\chi<0$).  This is previously known from the work of Monroe and Newman \cite{Monroe2005Impact} and later observed experimentally \cite{balsara2012-modadh, balsara2016-mod}.  For $v<1$, a previously unexplored regime in the context of electrodeposition, stability is achieved below the critical shear modulus ratio. The existence of density-driven stability may be understood in terms of the dependence of stability parameter $\chi$ on the hydrostatic term alone since the deviatoric term is always destabilizing. $\chi$ characterizes the electrochemical potential change of the electron at a peak in the interface ($\Delta \mu_{e^-}=A\chi$ when $\cos(kx)=1$). For $v<1$, the hydrostatic term in Eq. (\ref{eq:mu}) is stabilizing when $\Delta p^e + \Delta p^s$ is negative. Due to elongation of the electrode at a peak, there will be tensile stress generated at the electrode side of the interface and compressive at the electrolyte side. Hence $\Delta p^e<0$ and $\Delta p^s>0$. Since $G$ is a measure of the stress response to strain, when $G_s\ll G_e$, $|\Delta p^s|\ll |\Delta p^e|$ which will make this term stabilizing at low $G_s/G_e$. A similar argument explains the stable region on the top right. Thus, the stable regimes at the bottom left and top right in Fig. \ref{fig:phase-diag} are guaranteed to exist.

Based on the obtained criteria, we construct a stability diagram as shown in Fig. \ref{fig:phase-diag} with the shear modulus ratio and the molar volume ratio as the critical parameters. The electrode material used for generating the stability diagram is Li metal. The stability diagram has four regions out of which two are stable and two are unstable. The two stable regions lie on the top right and bottom left of the stability diagram. For $v>1$, a solid electrolyte with shear modulus larger than the critical shear modulus is required for stable electrodeposition.  In fact, the required shear modulus increases sharply as the molar volume ratio approaches unity, reducing the stability window.  The second region of stability emerges for $v<1$, which shows that it is possible to stabilize electrodeposition using a soft solid electrolyte provided Li in the solid electrolyte is more densely packed than Li in Li metal. We therefore term this stability mechanism as density-driven. Beyond $v=1$, stability requires the hydrostatic part of stress to dominate the stability parameter and hence, the stability in this region is called pressure-driven. This stability diagram qualitatively resembles the stability diagram for stress-driven phase transition at solid-solid interfaces studied by Angheluta et al. \cite{Angheluta2009, Angheluta2008}. In case of stress-driven phase transition, the interplay between the work term and elastic energy term determines the growth and stability of the interface. Analogously, it is the hydrostatic stress term, competing with the deviatoric stress term in electrodeposition.

This analysis raises the important question of where real solid electrolytes lie in this stability diagram.  This depends critically on the value of $v$ in solid electrolytes. Marcus and Hefter have tabulated the values of partial molar volumes of cations in a range of solvents \cite{marcus2004}. Following their work, for liquid and polymer electrolytes, the partial molar volume of the ion can be written as: $V=V_{\mathrm{int}}+V_{\mathrm{el}}+V_{\mathrm{cov}}+V_{\mathrm{str}}$, where the four terms correspond to intrinsic volume, and changes in the volume due to electrostriction, short-range interactions and size, shape and structure of solvent molecules. In crystalline solid electrolytes, the last three terms vanish and the partial molar volume is just the intrinsic volume of the ion in the crystal. We used the values of ionic radii tabulated by Shannon \cite{shannon1976revised} and Marcus et al. \cite{marcus-ionvolumes} (details in Supplemental Material \cite{Note2}). The values of the unit cell volume of solid electrolytes were obtained from \texttt{crystallography open database} \cite{COD2009} and the shear modulus from previous work on elastic properties of solid electrolytes \cite{Ahmad16uncertainty, Deng01012016} whenever available or from \texttt{materials project} database \cite{Jain2013, de2015charting}.

As shown in Fig. \ref{fig:phase-diag}, we find that typical inorganic solid electrolytes have a molar volume ratio, $v<1$ and possess a shear modulus higher than the critical shear modulus below which electrodeposition is stable.  As a result, Li-solid electrolyte interfaces based on these materials will result in unstable electrodeposition. Compounds in which Li has oxidation state of zero, like alloys of Li with Sn (not shown in Fig. \ref{fig:phase-diag}), generally have a molar volume ratio closer to 1. Solid polymer electrolytes generally have $v>1$ but their shear moduli are generally lower than the critical value, resulting in unstable electrodeposition.   Our analysis identifies a fundamental trade-off that needs to be broken if stable electrodeposition is expected for solid polymer or inorganic solid electrolytes. We note that the properties at the interface might change due to chemical reactions occurring at the reductive potentials of the anode. For example, different Li alloys might be formed at the interface depending on the composition of the solid electrolyte. In such cases, the effective properties at the interface must be used to determine stability.

Possible schemes for stable electrodeposition at metal-solid electrolyte interfaces rely on control of the shear modulus of the solid electrolyte or partial molar volume. An approach could be to alter the partial molar volume of Li in low shear modulus materials by tuning ion-solvent interactions so that they fall in the bottom left stable region on the stability diagram. Altering the shear modulus of the material is a much more difficult task requiring the use of strengthening mechanisms. Molten salts and ionic liquids with an elastic mechanical response that corresponds to low shear modulus could lie in the density-driven stability region. Finally, although the effect of defects like grain boundaries has not been included here, their effect may be included by determining the change in electrochemical potential of the components in Eq. (\ref{eq:reaction}). This will add a new term to $\Delta \mu_{e^-}$ in Eq. (\ref{eq:mu}).


In conclusion, we have explored the role of mechanics at solid-solid interfaces in determining electrodeposition stability. We show that two separate mechanisms of electrodeposition stability are possible: pressure-driven stability at high molar volume ratio and density-driven at lower molar volume ratio. These appear as two distinct regions in the stability diagram. Using these insights, we analyze candidate Li solid electrolytes, and show that materials re-engineering of the interface is required for stable electrodeposition.

\begin{acknowledgments}
Z. A. acknowledges support from the Advanced Research Projects Agency-Energy Integration and Optimization of Novel Ion Conducting Solids (IONICS) program under Grant No. \texttt{DE-AR0000774}. Z. A. and V. V. gratefully acknowledge support from the U.S. Department of Energy, Energy Efficiency and Renewable Energy Vehicle Technologies Office under Award No. \texttt{DE-EE0007810}.
\end{acknowledgments}

\bibliography{../../../solid-electrolytes/bibtexs/refs}

\providecommand{\noopsort}[1]{}\providecommand{\singleletter}[1]{#1}%
\begin{thebibliography}{46}%
\makeatletter
\providecommand \@ifxundefined [1]{%
 \@ifx{#1\undefined}
}%
\providecommand \@ifnum [1]{%
 \ifnum #1\expandafter \@firstoftwo
 \else \expandafter \@secondoftwo
 \fi
}%
\providecommand \@ifx [1]{%
 \ifx #1\expandafter \@firstoftwo
 \else \expandafter \@secondoftwo
 \fi
}%
\providecommand \natexlab [1]{#1}%
\providecommand \enquote  [1]{``#1''}%
\providecommand \bibnamefont  [1]{#1}%
\providecommand \bibfnamefont [1]{#1}%
\providecommand \citenamefont [1]{#1}%
\providecommand \href@noop [0]{\@secondoftwo}%
\providecommand \href [0]{\begingroup \@sanitize@url \@href}%
\providecommand \@href[1]{\@@startlink{#1}\@@href}%
\providecommand \@@href[1]{\endgroup#1\@@endlink}%
\providecommand \@sanitize@url [0]{\catcode `\\12\catcode `\$12\catcode
  `\&12\catcode `\#12\catcode `\^12\catcode `\_12\catcode `\%12\relax}%
\providecommand \@@startlink[1]{}%
\providecommand \@@endlink[0]{}%
\providecommand \url  [0]{\begingroup\@sanitize@url \@url }%
\providecommand \@url [1]{\endgroup\@href {#1}{\urlprefix }}%
\providecommand \urlprefix  [0]{URL }%
\providecommand \Eprint [0]{\href }%
\providecommand \doibase [0]{http://dx.doi.org/}%
\providecommand \selectlanguage [0]{\@gobble}%
\providecommand \bibinfo  [0]{\@secondoftwo}%
\providecommand \bibfield  [0]{\@secondoftwo}%
\providecommand \translation [1]{[#1]}%
\providecommand \BibitemOpen [0]{}%
\providecommand \bibitemStop [0]{}%
\providecommand \bibitemNoStop [0]{.\EOS\space}%
\providecommand \EOS [0]{\spacefactor3000\relax}%
\providecommand \BibitemShut  [1]{\csname bibitem#1\endcsname}%
\let\auto@bib@innerbib\@empty
\bibitem [{\citenamefont {Gamburg}\ and\ \citenamefont
  {Zangari}(2011)}]{gamburg2011theory}%
  \BibitemOpen
  \bibfield  {author} {\bibinfo {author} {\bibfnamefont {Y.~D.}\ \bibnamefont
  {Gamburg}}\ and\ \bibinfo {author} {\bibfnamefont {G.}~\bibnamefont
  {Zangari}},\ }\href@noop {} {\emph {\bibinfo {title} {Theory and Practice of
  Metal Electrodeposition}}}\ (\bibinfo  {publisher} {Springer, New York},\
  \bibinfo {address} {New York},\ \bibinfo {year} {2011})\BibitemShut {NoStop}%
\bibitem [{\citenamefont {Ross}(1994)}]{ross1994electrodeposited}%
  \BibitemOpen
  \bibfield  {author} {\bibinfo {author} {\bibfnamefont {C.}~\bibnamefont
  {Ross}},\ }\href@noop {} {\bibfield  {journal} {\bibinfo  {journal} {Annu.
  Rev. Mater. Sci.}\ }\textbf {\bibinfo {volume} {24}},\ \bibinfo {pages} {159}
  (\bibinfo {year} {1994})}\BibitemShut {NoStop}%
\bibitem [{\citenamefont {Barton}\ and\ \citenamefont
  {Bockris}(1962)}]{Barton1962}%
  \BibitemOpen
  \bibfield  {author} {\bibinfo {author} {\bibfnamefont {J.~L.}\ \bibnamefont
  {Barton}}\ and\ \bibinfo {author} {\bibfnamefont {J.~O.}\ \bibnamefont
  {Bockris}},\ }\href@noop {} {\bibfield  {journal} {\bibinfo  {journal} {Proc.
  R. Soc. A}\ }\textbf {\bibinfo {volume} {268}},\ \bibinfo {pages} {485}
  (\bibinfo {year} {1962})}\BibitemShut {NoStop}%
\bibitem [{\citenamefont {Chazalviel}(1990)}]{Chazalviel90}%
  \BibitemOpen
  \bibfield  {author} {\bibinfo {author} {\bibfnamefont {J.-N.}\ \bibnamefont
  {Chazalviel}},\ }\href@noop {} {\bibfield  {journal} {\bibinfo  {journal}
  {Phys. Rev. A}\ }\textbf {\bibinfo {volume} {42}},\ \bibinfo {pages} {7355}
  (\bibinfo {year} {1990})}\BibitemShut {NoStop}%
\bibitem [{\citenamefont {Shraiman}\ and\ \citenamefont
  {Bensimon}(1984)}]{Shraiman84}%
  \BibitemOpen
  \bibfield  {author} {\bibinfo {author} {\bibfnamefont {B.}~\bibnamefont
  {Shraiman}}\ and\ \bibinfo {author} {\bibfnamefont {D.}~\bibnamefont
  {Bensimon}},\ }\href@noop {} {\bibfield  {journal} {\bibinfo  {journal}
  {Phys. Rev. A}\ }\textbf {\bibinfo {volume} {30}},\ \bibinfo {pages} {2840}
  (\bibinfo {year} {1984})}\BibitemShut {NoStop}%
\bibitem [{\citenamefont {Voss}\ and\ \citenamefont
  {Tomkiewicz}(1985)}]{Voss85}%
  \BibitemOpen
  \bibfield  {author} {\bibinfo {author} {\bibfnamefont {R.~F.}\ \bibnamefont
  {Voss}}\ and\ \bibinfo {author} {\bibfnamefont {M.}~\bibnamefont
  {Tomkiewicz}},\ }\href@noop {} {\bibfield  {journal} {\bibinfo  {journal} {J.
  Electrochem. Soc.}\ }\textbf {\bibinfo {volume} {132}},\ \bibinfo {pages}
  {371} (\bibinfo {year} {1985})}\BibitemShut {NoStop}%
\bibitem [{\citenamefont {Sawada}\ \emph {et~al.}(1986)\citenamefont {Sawada},
  \citenamefont {Dougherty},\ and\ \citenamefont
  {Gollub}}]{PhysRevLett.56.1260}%
  \BibitemOpen
  \bibfield  {author} {\bibinfo {author} {\bibfnamefont {Y.}~\bibnamefont
  {Sawada}}, \bibinfo {author} {\bibfnamefont {A.}~\bibnamefont {Dougherty}}, \
  and\ \bibinfo {author} {\bibfnamefont {J.~P.}\ \bibnamefont {Gollub}},\
  }\href@noop {} {\bibfield  {journal} {\bibinfo  {journal} {Phys. Rev. Lett.}\
  }\textbf {\bibinfo {volume} {56}},\ \bibinfo {pages} {1260} (\bibinfo {year}
  {1986})}\BibitemShut {NoStop}%
\bibitem [{\citenamefont {L\`opez-Salvans}\ \emph {et~al.}(1996)\citenamefont
  {L\`opez-Salvans}, \citenamefont {Trigueros}, \citenamefont {Vallmitjana},
  \citenamefont {Claret},\ and\ \citenamefont
  {Sagu\'es}}]{PhysRevLett.76.4062}%
  \BibitemOpen
  \bibfield  {author} {\bibinfo {author} {\bibfnamefont {M.-Q.}\ \bibnamefont
  {L\`opez-Salvans}}, \bibinfo {author} {\bibfnamefont {P.~P.}\ \bibnamefont
  {Trigueros}}, \bibinfo {author} {\bibfnamefont {S.}~\bibnamefont
  {Vallmitjana}}, \bibinfo {author} {\bibfnamefont {J.}~\bibnamefont {Claret}},
  \ and\ \bibinfo {author} {\bibfnamefont {F.}~\bibnamefont {Sagu\'es}},\
  }\href@noop {} {\bibfield  {journal} {\bibinfo  {journal} {Phys. Rev. Lett.}\
  }\textbf {\bibinfo {volume} {76}},\ \bibinfo {pages} {4062} (\bibinfo {year}
  {1996})}\BibitemShut {NoStop}%
\bibitem [{\citenamefont {Argoul}\ \emph {et~al.}(1988)\citenamefont {Argoul},
  \citenamefont {Arneodo}, \citenamefont {Grasseau},\ and\ \citenamefont
  {Swinney}}]{PhysRevLett.61.2558}%
  \BibitemOpen
  \bibfield  {author} {\bibinfo {author} {\bibfnamefont {F.}~\bibnamefont
  {Argoul}}, \bibinfo {author} {\bibfnamefont {A.}~\bibnamefont {Arneodo}},
  \bibinfo {author} {\bibfnamefont {G.}~\bibnamefont {Grasseau}}, \ and\
  \bibinfo {author} {\bibfnamefont {H.~L.}\ \bibnamefont {Swinney}},\
  }\href@noop {} {\bibfield  {journal} {\bibinfo  {journal} {Phys. Rev. Lett.}\
  }\textbf {\bibinfo {volume} {61}},\ \bibinfo {pages} {2558} (\bibinfo {year}
  {1988})}\BibitemShut {NoStop}%
\bibitem [{\citenamefont {Huo}\ and\ \citenamefont
  {Schwarzacher}(2001)}]{PhysRevLett.86.256}%
  \BibitemOpen
  \bibfield  {author} {\bibinfo {author} {\bibfnamefont {S.}~\bibnamefont
  {Huo}}\ and\ \bibinfo {author} {\bibfnamefont {W.}~\bibnamefont
  {Schwarzacher}},\ }\href@noop {} {\bibfield  {journal} {\bibinfo  {journal}
  {Phys. Rev. Lett.}\ }\textbf {\bibinfo {volume} {86}},\ \bibinfo {pages}
  {256} (\bibinfo {year} {2001})}\BibitemShut {NoStop}%
\bibitem [{\citenamefont {Matsushita}\ \emph {et~al.}(1984)\citenamefont
  {Matsushita}, \citenamefont {Sano}, \citenamefont {Hayakawa}, \citenamefont
  {Honjo},\ and\ \citenamefont {Sawada}}]{MatsushitaFractal1984}%
  \BibitemOpen
  \bibfield  {author} {\bibinfo {author} {\bibfnamefont {M.}~\bibnamefont
  {Matsushita}}, \bibinfo {author} {\bibfnamefont {M.}~\bibnamefont {Sano}},
  \bibinfo {author} {\bibfnamefont {Y.}~\bibnamefont {Hayakawa}}, \bibinfo
  {author} {\bibfnamefont {H.}~\bibnamefont {Honjo}}, \ and\ \bibinfo {author}
  {\bibfnamefont {Y.}~\bibnamefont {Sawada}},\ }\href@noop {} {\bibfield
  {journal} {\bibinfo  {journal} {Phys. Rev. Lett.}\ }\textbf {\bibinfo
  {volume} {53}},\ \bibinfo {pages} {286} (\bibinfo {year} {1984})}\BibitemShut
  {NoStop}%
\bibitem [{\citenamefont {Erlebacher}\ \emph {et~al.}(1993)\citenamefont
  {Erlebacher}, \citenamefont {Searson},\ and\ \citenamefont
  {Sieradzki}}]{erlebacher-93-comp}%
  \BibitemOpen
  \bibfield  {author} {\bibinfo {author} {\bibfnamefont {J.}~\bibnamefont
  {Erlebacher}}, \bibinfo {author} {\bibfnamefont {P.~C.}\ \bibnamefont
  {Searson}}, \ and\ \bibinfo {author} {\bibfnamefont {K.}~\bibnamefont
  {Sieradzki}},\ }\href@noop {} {\bibfield  {journal} {\bibinfo  {journal}
  {Phys. Rev. Lett.}\ }\textbf {\bibinfo {volume} {71}},\ \bibinfo {pages}
  {3311} (\bibinfo {year} {1993})}\BibitemShut {NoStop}%
\bibitem [{\citenamefont {Aurbach}\ \emph {et~al.}(2002)\citenamefont
  {Aurbach}, \citenamefont {Zinigrad}, \citenamefont {Cohen},\ and\
  \citenamefont {Teller}}]{aurbach2002short}%
  \BibitemOpen
  \bibfield  {author} {\bibinfo {author} {\bibfnamefont {D.}~\bibnamefont
  {Aurbach}}, \bibinfo {author} {\bibfnamefont {E.}~\bibnamefont {Zinigrad}},
  \bibinfo {author} {\bibfnamefont {Y.}~\bibnamefont {Cohen}}, \ and\ \bibinfo
  {author} {\bibfnamefont {H.}~\bibnamefont {Teller}},\ }\href@noop {}
  {\bibfield  {journal} {\bibinfo  {journal} {Solid State Ionics}\ }\textbf
  {\bibinfo {volume} {148}},\ \bibinfo {pages} {405} (\bibinfo {year}
  {2002})}\BibitemShut {NoStop}%
\bibitem [{\citenamefont {Xu}\ \emph {et~al.}(2014)\citenamefont {Xu},
  \citenamefont {Wang}, \citenamefont {Ding}, \citenamefont {Chen},
  \citenamefont {Nasybulin}, \citenamefont {Zhang},\ and\ \citenamefont
  {Zhang}}]{XuLi2014}%
  \BibitemOpen
  \bibfield  {author} {\bibinfo {author} {\bibfnamefont {W.}~\bibnamefont
  {Xu}}, \bibinfo {author} {\bibfnamefont {J.}~\bibnamefont {Wang}}, \bibinfo
  {author} {\bibfnamefont {F.}~\bibnamefont {Ding}}, \bibinfo {author}
  {\bibfnamefont {X.}~\bibnamefont {Chen}}, \bibinfo {author} {\bibfnamefont
  {E.}~\bibnamefont {Nasybulin}}, \bibinfo {author} {\bibfnamefont
  {Y.}~\bibnamefont {Zhang}}, \ and\ \bibinfo {author} {\bibfnamefont {J.-G.}\
  \bibnamefont {Zhang}},\ }\href@noop {} {\bibfield  {journal} {\bibinfo
  {journal} {Energy Environ. Sci.}\ }\textbf {\bibinfo {volume} {7}},\ \bibinfo
  {pages} {513} (\bibinfo {year} {2014})}\BibitemShut {NoStop}%
\bibitem [{\citenamefont {Sapunkov}\ \emph {et~al.}(2015)\citenamefont
  {Sapunkov}, \citenamefont {Pande}, \citenamefont {Khetan}, \citenamefont
  {Choomwattana},\ and\ \citenamefont {Viswanathan}}]{oleg2015TMR}%
  \BibitemOpen
  \bibfield  {author} {\bibinfo {author} {\bibfnamefont {O.}~\bibnamefont
  {Sapunkov}}, \bibinfo {author} {\bibfnamefont {V.}~\bibnamefont {Pande}},
  \bibinfo {author} {\bibfnamefont {A.}~\bibnamefont {Khetan}}, \bibinfo
  {author} {\bibfnamefont {C.}~\bibnamefont {Choomwattana}}, \ and\ \bibinfo
  {author} {\bibfnamefont {V.}~\bibnamefont {Viswanathan}},\ }\href@noop {}
  {\bibfield  {journal} {\bibinfo  {journal} {Transl. Mater. Res.}\ }\textbf
  {\bibinfo {volume} {2}},\ \bibinfo {pages} {045002} (\bibinfo {year}
  {2015})}\BibitemShut {NoStop}%
\bibitem [{\citenamefont {Schaefer}\ \emph {et~al.}(2012)\citenamefont
  {Schaefer}, \citenamefont {Lu}, \citenamefont {Moganty}, \citenamefont
  {Agarwal}, \citenamefont {Jayaprakash},\ and\ \citenamefont
  {Archer}}]{Schaefer2012}%
  \BibitemOpen
  \bibfield  {author} {\bibinfo {author} {\bibfnamefont {J.~L.}\ \bibnamefont
  {Schaefer}}, \bibinfo {author} {\bibfnamefont {Y.}~\bibnamefont {Lu}},
  \bibinfo {author} {\bibfnamefont {S.~S.}\ \bibnamefont {Moganty}}, \bibinfo
  {author} {\bibfnamefont {P.}~\bibnamefont {Agarwal}}, \bibinfo {author}
  {\bibfnamefont {N.}~\bibnamefont {Jayaprakash}}, \ and\ \bibinfo {author}
  {\bibfnamefont {L.~A.}\ \bibnamefont {Archer}},\ }\href@noop {} {\bibfield
  {journal} {\bibinfo  {journal} {Appl. Nanosci.}\ }\textbf {\bibinfo {volume}
  {2}},\ \bibinfo {pages} {91} (\bibinfo {year} {2012})}\BibitemShut {NoStop}%
\bibitem [{\citenamefont {Tikekar}\ \emph
  {et~al.}(2016{\natexlab{a}})\citenamefont {Tikekar}, \citenamefont
  {Choudhury}, \citenamefont {Tu},\ and\ \citenamefont
  {Archer}}]{TikekarNatEnergy2016}%
  \BibitemOpen
  \bibfield  {author} {\bibinfo {author} {\bibfnamefont {M.~D.}\ \bibnamefont
  {Tikekar}}, \bibinfo {author} {\bibfnamefont {S.}~\bibnamefont {Choudhury}},
  \bibinfo {author} {\bibfnamefont {Z.}~\bibnamefont {Tu}}, \ and\ \bibinfo
  {author} {\bibfnamefont {L.~A.}\ \bibnamefont {Archer}},\ }\href@noop {}
  {\bibfield  {journal} {\bibinfo  {journal} {Nat. Energy}\ }\textbf {\bibinfo
  {volume} {1}},\ \bibinfo {pages} {16114} (\bibinfo {year}
  {2016}{\natexlab{a}})}\BibitemShut {NoStop}%
\bibitem [{\citenamefont {Monroe}\ and\ \citenamefont
  {Newman}(2004)}]{Monroe2004Effect}%
  \BibitemOpen
  \bibfield  {author} {\bibinfo {author} {\bibfnamefont {C.}~\bibnamefont
  {Monroe}}\ and\ \bibinfo {author} {\bibfnamefont {J.}~\bibnamefont
  {Newman}},\ }\href@noop {} {\bibfield  {journal} {\bibinfo  {journal} {J.
  Electrochem. Soc.}\ }\textbf {\bibinfo {volume} {151}},\ \bibinfo {pages}
  {A880} (\bibinfo {year} {2004})}\BibitemShut {NoStop}%
\bibitem [{\citenamefont {Monroe}\ and\ \citenamefont
  {Newman}(2005)}]{Monroe2005Impact}%
  \BibitemOpen
  \bibfield  {author} {\bibinfo {author} {\bibfnamefont {C.}~\bibnamefont
  {Monroe}}\ and\ \bibinfo {author} {\bibfnamefont {J.}~\bibnamefont
  {Newman}},\ }\href@noop {} {\bibfield  {journal} {\bibinfo  {journal} {J.
  Electrochem. Soc.}\ }\textbf {\bibinfo {volume} {152}},\ \bibinfo {pages}
  {A396} (\bibinfo {year} {2005})}\BibitemShut {NoStop}%
\bibitem [{\citenamefont {Angheluta}\ \emph {et~al.}(2009)\citenamefont
  {Angheluta}, \citenamefont {Jettestuen},\ and\ \citenamefont
  {Mathiesen}}]{Angheluta2009}%
  \BibitemOpen
  \bibfield  {author} {\bibinfo {author} {\bibfnamefont {L.}~\bibnamefont
  {Angheluta}}, \bibinfo {author} {\bibfnamefont {E.}~\bibnamefont
  {Jettestuen}}, \ and\ \bibinfo {author} {\bibfnamefont {J.}~\bibnamefont
  {Mathiesen}},\ }\href@noop {} {\bibfield  {journal} {\bibinfo  {journal}
  {Phys. Rev. E}\ }\textbf {\bibinfo {volume} {79}},\ \bibinfo {pages} {031601}
  (\bibinfo {year} {2009})}\BibitemShut {NoStop}%
\bibitem [{Note1()}]{Note1}%
  \BibitemOpen
  \bibinfo {note} {The initiation regime has been shown to be most critical for
  dendrite suppression since dendrites cannot be suppressed if they reach
  propagation regime. See Ref. \protect \rev@citealp {Monroe2004Effect} and
  references cited in its introduction.}\BibitemShut {Stop}%
\bibitem [{\citenamefont {Ren}\ \emph {et~al.}(2015)\citenamefont {Ren},
  \citenamefont {Shen}, \citenamefont {Lin},\ and\ \citenamefont
  {Nan}}]{Ren2015direct}%
  \BibitemOpen
  \bibfield  {author} {\bibinfo {author} {\bibfnamefont {Y.}~\bibnamefont
  {Ren}}, \bibinfo {author} {\bibfnamefont {Y.}~\bibnamefont {Shen}}, \bibinfo
  {author} {\bibfnamefont {Y.}~\bibnamefont {Lin}}, \ and\ \bibinfo {author}
  {\bibfnamefont {C.-W.}\ \bibnamefont {Nan}},\ }\href@noop {} {\bibfield
  {journal} {\bibinfo  {journal} {Electrochem. Commun.}\ }\textbf {\bibinfo
  {volume} {57}},\ \bibinfo {pages} {27 } (\bibinfo {year} {2015})}\BibitemShut
  {NoStop}%
\bibitem [{\citenamefont {Newman}\ and\ \citenamefont
  {Thomas-Alyea}(2012)}]{newman2012electrochemical}%
  \BibitemOpen
  \bibfield  {author} {\bibinfo {author} {\bibfnamefont {J.}~\bibnamefont
  {Newman}}\ and\ \bibinfo {author} {\bibfnamefont {K.~E.}\ \bibnamefont
  {Thomas-Alyea}},\ }\href@noop {} {\emph {\bibinfo {title} {Electrochemical
  Systems}}}\ (\bibinfo  {publisher} {John Wiley \& Sons},\ \bibinfo {address}
  {Hoboken},\ \bibinfo {year} {2012})\BibitemShut {NoStop}%
\bibitem [{\citenamefont {Bazant}\ \emph {et~al.}(2005)\citenamefont {Bazant},
  \citenamefont {Chu},\ and\ \citenamefont {Bayly}}]{bazant2005}%
  \BibitemOpen
  \bibfield  {author} {\bibinfo {author} {\bibfnamefont {M.~Z.}\ \bibnamefont
  {Bazant}}, \bibinfo {author} {\bibfnamefont {K.~T.}\ \bibnamefont {Chu}}, \
  and\ \bibinfo {author} {\bibfnamefont {B.~J.}\ \bibnamefont {Bayly}},\
  }\href@noop {} {\bibfield  {journal} {\bibinfo  {journal} {SIAM J. Appl.
  Math.}\ }\textbf {\bibinfo {volume} {65}},\ \bibinfo {pages} {1463} (\bibinfo
  {year} {2005})}\BibitemShut {NoStop}%
\bibitem [{\citenamefont {Mullins}\ and\ \citenamefont
  {Sekerka}(1963)}]{MullinsSekerka63}%
  \BibitemOpen
  \bibfield  {author} {\bibinfo {author} {\bibfnamefont {W.~W.}\ \bibnamefont
  {Mullins}}\ and\ \bibinfo {author} {\bibfnamefont {R.~F.}\ \bibnamefont
  {Sekerka}},\ }\href@noop {} {\bibfield  {journal} {\bibinfo  {journal} {J.
  Appl. Phys.}\ }\textbf {\bibinfo {volume} {34}},\ \bibinfo {pages} {323}
  (\bibinfo {year} {1963})}\BibitemShut {NoStop}%
\bibitem [{\citenamefont {Mullins}\ and\ \citenamefont
  {Sekerka}(1964)}]{MullinsSekerka64}%
  \BibitemOpen
  \bibfield  {author} {\bibinfo {author} {\bibfnamefont {W.~W.}\ \bibnamefont
  {Mullins}}\ and\ \bibinfo {author} {\bibfnamefont {R.~F.}\ \bibnamefont
  {Sekerka}},\ }\href@noop {} {\bibfield  {journal} {\bibinfo  {journal} {J.
  Appl. Phys.}\ }\textbf {\bibinfo {volume} {35}},\ \bibinfo {pages} {444}
  (\bibinfo {year} {1964})}\BibitemShut {NoStop}%
\bibitem [{\citenamefont {Carter}\ and\ \citenamefont
  {Handwerker}(1993)}]{Carter93}%
  \BibitemOpen
  \bibfield  {author} {\bibinfo {author} {\bibfnamefont {W.}~\bibnamefont
  {Carter}}\ and\ \bibinfo {author} {\bibfnamefont {C.}~\bibnamefont
  {Handwerker}},\ }\href@noop {} {\bibfield  {journal} {\bibinfo  {journal}
  {Acta Metall. Mater.}\ }\textbf {\bibinfo {volume} {41}},\ \bibinfo {pages}
  {1633 } (\bibinfo {year} {1993})}\BibitemShut {NoStop}%
\bibitem [{\citenamefont {Tikekar}\ \emph
  {et~al.}(2016{\natexlab{b}})\citenamefont {Tikekar}, \citenamefont {Archer},\
  and\ \citenamefont {Koch}}]{Tikekar2016}%
  \BibitemOpen
  \bibfield  {author} {\bibinfo {author} {\bibfnamefont {M.~D.}\ \bibnamefont
  {Tikekar}}, \bibinfo {author} {\bibfnamefont {L.~A.}\ \bibnamefont {Archer}},
  \ and\ \bibinfo {author} {\bibfnamefont {D.~L.}\ \bibnamefont {Koch}},\
  }\href@noop {} {\bibfield  {journal} {\bibinfo  {journal} {Sci. Adv.}\
  }\textbf {\bibinfo {volume} {2}},\ \bibinfo {pages} {1600320} (\bibinfo
  {year} {2016}{\natexlab{b}})}\BibitemShut {NoStop}%
\bibitem [{\citenamefont {Asaro}\ and\ \citenamefont
  {Tiller}(1972)}]{Asaro1972}%
  \BibitemOpen
  \bibfield  {author} {\bibinfo {author} {\bibfnamefont {R.~J.}\ \bibnamefont
  {Asaro}}\ and\ \bibinfo {author} {\bibfnamefont {W.~A.}\ \bibnamefont
  {Tiller}},\ }\href@noop {} {\bibfield  {journal} {\bibinfo  {journal}
  {Metall. Trans.}\ }\textbf {\bibinfo {volume} {3}},\ \bibinfo {pages} {1789}
  (\bibinfo {year} {1972})}\BibitemShut {NoStop}%
\bibitem [{\citenamefont {Bullard}\ \emph {et~al.}(1998)\citenamefont
  {Bullard}, \citenamefont {Garboczi},\ and\ \citenamefont
  {Carter}}]{carter-asaro-1998}%
  \BibitemOpen
  \bibfield  {author} {\bibinfo {author} {\bibfnamefont {J.~W.}\ \bibnamefont
  {Bullard}}, \bibinfo {author} {\bibfnamefont {E.~J.}\ \bibnamefont
  {Garboczi}}, \ and\ \bibinfo {author} {\bibfnamefont {W.~C.}\ \bibnamefont
  {Carter}},\ }\href@noop {} {\bibfield  {journal} {\bibinfo  {journal} {J.
  Appl. Phys.}\ }\textbf {\bibinfo {volume} {83}},\ \bibinfo {pages} {4477}
  (\bibinfo {year} {1998})}\BibitemShut {NoStop}%
\bibitem [{Note2()}]{Note2}%
  \BibitemOpen
  \bibinfo {note} {See Supplemental Material for exact expression for $\chi $
  and critical shear modulus ratio, and calculations of partial molar volume
  ratio.}\BibitemShut {Stop}%
\bibitem [{\citenamefont {Freund}\ and\ \citenamefont
  {Suresh}(2004)}]{freund2004thin}%
  \BibitemOpen
  \bibfield  {author} {\bibinfo {author} {\bibfnamefont {L.~B.}\ \bibnamefont
  {Freund}}\ and\ \bibinfo {author} {\bibfnamefont {S.}~\bibnamefont
  {Suresh}},\ }\href@noop {} {\emph {\bibinfo {title} {Thin Film Materials:
  Stress, Defect Formation and Surface Evolution}}}\ (\bibinfo  {publisher}
  {Cambridge University Press},\ \bibinfo {address} {Cambridge, England},\
  \bibinfo {year} {2004})\BibitemShut {NoStop}%
\bibitem [{\citenamefont {Marcus}\ and\ \citenamefont
  {Hefter}(2004)}]{marcus2004}%
  \BibitemOpen
  \bibfield  {author} {\bibinfo {author} {\bibfnamefont {Y.}~\bibnamefont
  {Marcus}}\ and\ \bibinfo {author} {\bibfnamefont {G.}~\bibnamefont
  {Hefter}},\ }\href@noop {} {\bibfield  {journal} {\bibinfo  {journal} {Chem.
  Rev.}\ }\textbf {\bibinfo {volume} {104}},\ \bibinfo {pages} {3405} (\bibinfo
  {year} {2004})}\BibitemShut {NoStop}%
\bibitem [{\citenamefont {Wang}\ \emph {et~al.}(2017)\citenamefont {Wang},
  \citenamefont {Zhang}, \citenamefont {Zheng}, \citenamefont {Cui},
  \citenamefont {Rojo},\ and\ \citenamefont {Zhang}}]{Wang17}%
  \BibitemOpen
  \bibfield  {author} {\bibinfo {author} {\bibfnamefont {D.}~\bibnamefont
  {Wang}}, \bibinfo {author} {\bibfnamefont {W.}~\bibnamefont {Zhang}},
  \bibinfo {author} {\bibfnamefont {W.}~\bibnamefont {Zheng}}, \bibinfo
  {author} {\bibfnamefont {X.}~\bibnamefont {Cui}}, \bibinfo {author}
  {\bibfnamefont {T.}~\bibnamefont {Rojo}}, \ and\ \bibinfo {author}
  {\bibfnamefont {Q.}~\bibnamefont {Zhang}},\ }\href@noop {} {\bibfield
  {journal} {\bibinfo  {journal} {Adv. Sci.}\ }\textbf {\bibinfo {volume}
  {4}},\ \bibinfo {pages} {1600168} (\bibinfo {year} {2017})}\BibitemShut
  {NoStop}%
\bibitem [{\citenamefont {Vitos}\ \emph {et~al.}(1998)\citenamefont {Vitos},
  \citenamefont {Ruban}, \citenamefont {Skriver},\ and\ \citenamefont
  {Koll{\'a}r}}]{Vitos98surface}%
  \BibitemOpen
  \bibfield  {author} {\bibinfo {author} {\bibfnamefont {L.}~\bibnamefont
  {Vitos}}, \bibinfo {author} {\bibfnamefont {A.}~\bibnamefont {Ruban}},
  \bibinfo {author} {\bibfnamefont {H.}~\bibnamefont {Skriver}}, \ and\
  \bibinfo {author} {\bibfnamefont {J.}~\bibnamefont {Koll{\'a}r}},\
  }\href@noop {} {\bibfield  {journal} {\bibinfo  {journal} {Surf. Sci.}\
  }\textbf {\bibinfo {volume} {411}},\ \bibinfo {pages} {186 } (\bibinfo {year}
  {1998})}\BibitemShut {NoStop}%
\bibitem [{\citenamefont {Stone}\ \emph {et~al.}(2012)\citenamefont {Stone},
  \citenamefont {Mullin}, \citenamefont {Teran}, \citenamefont {Hallinan},
  \citenamefont {Minor}, \citenamefont {Hexemer},\ and\ \citenamefont
  {Balsara}}]{balsara2012-modadh}%
  \BibitemOpen
  \bibfield  {author} {\bibinfo {author} {\bibfnamefont {G.~M.}\ \bibnamefont
  {Stone}}, \bibinfo {author} {\bibfnamefont {S.~A.}\ \bibnamefont {Mullin}},
  \bibinfo {author} {\bibfnamefont {A.~A.}\ \bibnamefont {Teran}}, \bibinfo
  {author} {\bibfnamefont {D.~T.}\ \bibnamefont {Hallinan}}, \bibinfo {author}
  {\bibfnamefont {A.~M.}\ \bibnamefont {Minor}}, \bibinfo {author}
  {\bibfnamefont {A.}~\bibnamefont {Hexemer}}, \ and\ \bibinfo {author}
  {\bibfnamefont {N.~P.}\ \bibnamefont {Balsara}},\ }\href@noop {} {\bibfield
  {journal} {\bibinfo  {journal} {J. Electrochem. Soc.}\ }\textbf {\bibinfo
  {volume} {159}},\ \bibinfo {pages} {A222} (\bibinfo {year}
  {2012})}\BibitemShut {NoStop}%
\bibitem [{\citenamefont {Harry}\ \emph {et~al.}(2016)\citenamefont {Harry},
  \citenamefont {Higa}, \citenamefont {Srinivasan},\ and\ \citenamefont
  {Balsara}}]{balsara2016-mod}%
  \BibitemOpen
  \bibfield  {author} {\bibinfo {author} {\bibfnamefont {K.~J.}\ \bibnamefont
  {Harry}}, \bibinfo {author} {\bibfnamefont {K.}~\bibnamefont {Higa}},
  \bibinfo {author} {\bibfnamefont {V.}~\bibnamefont {Srinivasan}}, \ and\
  \bibinfo {author} {\bibfnamefont {N.~P.}\ \bibnamefont {Balsara}},\
  }\href@noop {} {\bibfield  {journal} {\bibinfo  {journal} {J. Electrochem.
  Soc.}\ }\textbf {\bibinfo {volume} {163}},\ \bibinfo {pages} {A2216}
  (\bibinfo {year} {2016})}\BibitemShut {NoStop}%
\bibitem [{\citenamefont {Angheluta}\ \emph {et~al.}(2008)\citenamefont
  {Angheluta}, \citenamefont {Jettestuen}, \citenamefont {Mathiesen},
  \citenamefont {Renard},\ and\ \citenamefont {Jamtveit}}]{Angheluta2008}%
  \BibitemOpen
  \bibfield  {author} {\bibinfo {author} {\bibfnamefont {L.}~\bibnamefont
  {Angheluta}}, \bibinfo {author} {\bibfnamefont {E.}~\bibnamefont
  {Jettestuen}}, \bibinfo {author} {\bibfnamefont {J.}~\bibnamefont
  {Mathiesen}}, \bibinfo {author} {\bibfnamefont {F.}~\bibnamefont {Renard}}, \
  and\ \bibinfo {author} {\bibfnamefont {B.}~\bibnamefont {Jamtveit}},\
  }\href@noop {} {\bibfield  {journal} {\bibinfo  {journal} {Phys. Rev. Lett.}\
  }\textbf {\bibinfo {volume} {100}},\ \bibinfo {pages} {096105} (\bibinfo
  {year} {2008})}\BibitemShut {NoStop}%
\bibitem [{\citenamefont {Hafezi}(2002)}]{Hafezi2002}%
  \BibitemOpen
  \bibfield  {author} {\bibinfo {author} {\bibfnamefont {H.}~\bibnamefont
  {Hafezi}},\ }\href@noop {} {\bibinfo {type} {{Ph.D.} thesis}},\ \bibinfo
  {school} {University of California Berkeley} (\bibinfo {year}
  {2002})\BibitemShut {NoStop}%
\bibitem [{\citenamefont {Shannon}(1976)}]{shannon1976revised}%
  \BibitemOpen
  \bibfield  {author} {\bibinfo {author} {\bibfnamefont {R.~D.}\ \bibnamefont
  {Shannon}},\ }\href@noop {} {\bibfield  {journal} {\bibinfo  {journal} {Acta
  Crystallogr., Sect. A}\ }\textbf {\bibinfo {volume} {32}},\ \bibinfo {pages}
  {751} (\bibinfo {year} {1976})}\BibitemShut {NoStop}%
\bibitem [{\citenamefont {Marcus}\ \emph {et~al.}(2002)\citenamefont {Marcus},
  \citenamefont {Donald Brooke~Jenkins},\ and\ \citenamefont
  {Glasser}}]{marcus-ionvolumes}%
  \BibitemOpen
  \bibfield  {author} {\bibinfo {author} {\bibfnamefont {Y.}~\bibnamefont
  {Marcus}}, \bibinfo {author} {\bibfnamefont {H.}~\bibnamefont {Donald
  Brooke~Jenkins}}, \ and\ \bibinfo {author} {\bibfnamefont {L.}~\bibnamefont
  {Glasser}},\ }\href@noop {} {\bibfield  {journal} {\bibinfo  {journal} {J.
  Chem. Soc., Dalton Trans.}\ }\textbf {\bibinfo {volume} {2002}},\ \bibinfo
  {pages} {3795} (\bibinfo {year} {2002})}\BibitemShut {NoStop}%
\bibitem [{\citenamefont {Gra{\v{z}}ulis}\ \emph {et~al.}(2009)\citenamefont
  {Gra{\v{z}}ulis}, \citenamefont {Chateigner}, \citenamefont {Downs},
  \citenamefont {Yokochi}, \citenamefont {Quir{\'o}s}, \citenamefont
  {Lutterotti}, \citenamefont {Manakova}, \citenamefont {Butkus}, \citenamefont
  {Moeck},\ and\ \citenamefont {Le~Bail}}]{COD2009}%
  \BibitemOpen
  \bibfield  {author} {\bibinfo {author} {\bibfnamefont {S.}~\bibnamefont
  {Gra{\v{z}}ulis}}, \bibinfo {author} {\bibfnamefont {D.}~\bibnamefont
  {Chateigner}}, \bibinfo {author} {\bibfnamefont {R.~T.}\ \bibnamefont
  {Downs}}, \bibinfo {author} {\bibfnamefont {A.}~\bibnamefont {Yokochi}},
  \bibinfo {author} {\bibfnamefont {M.}~\bibnamefont {Quir{\'o}s}}, \bibinfo
  {author} {\bibfnamefont {L.}~\bibnamefont {Lutterotti}}, \bibinfo {author}
  {\bibfnamefont {E.}~\bibnamefont {Manakova}}, \bibinfo {author}
  {\bibfnamefont {J.}~\bibnamefont {Butkus}}, \bibinfo {author} {\bibfnamefont
  {P.}~\bibnamefont {Moeck}}, \ and\ \bibinfo {author} {\bibfnamefont
  {A.}~\bibnamefont {Le~Bail}},\ }\href@noop {} {\bibfield  {journal} {\bibinfo
   {journal} {J. Appl. Crystallogr.}\ }\textbf {\bibinfo {volume} {42}},\
  \bibinfo {pages} {726} (\bibinfo {year} {2009})}\BibitemShut {NoStop}%
\bibitem [{\citenamefont {Ahmad}\ and\ \citenamefont
  {Viswanathan}(2016)}]{Ahmad16uncertainty}%
  \BibitemOpen
  \bibfield  {author} {\bibinfo {author} {\bibfnamefont {Z.}~\bibnamefont
  {Ahmad}}\ and\ \bibinfo {author} {\bibfnamefont {V.}~\bibnamefont
  {Viswanathan}},\ }\href@noop {} {\bibfield  {journal} {\bibinfo  {journal}
  {Phys. Rev. B}\ }\textbf {\bibinfo {volume} {94}},\ \bibinfo {pages} {064105}
  (\bibinfo {year} {2016})}\BibitemShut {NoStop}%
\bibitem [{\citenamefont {Deng}\ \emph {et~al.}(2016)\citenamefont {Deng},
  \citenamefont {Wang}, \citenamefont {Chu}, \citenamefont {Luo},\ and\
  \citenamefont {Ong}}]{Deng01012016}%
  \BibitemOpen
  \bibfield  {author} {\bibinfo {author} {\bibfnamefont {Z.}~\bibnamefont
  {Deng}}, \bibinfo {author} {\bibfnamefont {Z.}~\bibnamefont {Wang}}, \bibinfo
  {author} {\bibfnamefont {I.-H.}\ \bibnamefont {Chu}}, \bibinfo {author}
  {\bibfnamefont {J.}~\bibnamefont {Luo}}, \ and\ \bibinfo {author}
  {\bibfnamefont {S.~P.}\ \bibnamefont {Ong}},\ }\href@noop {} {\bibfield
  {journal} {\bibinfo  {journal} {J. Electrochem. Soc.}\ }\textbf {\bibinfo
  {volume} {163}},\ \bibinfo {pages} {A67} (\bibinfo {year}
  {2016})}\BibitemShut {NoStop}%
\bibitem [{\citenamefont {Jain}\ \emph {et~al.}(2013)\citenamefont {Jain},
  \citenamefont {Ong}, \citenamefont {Hautier}, \citenamefont {Chen},
  \citenamefont {Richards}, \citenamefont {Dacek}, \citenamefont {Cholia},
  \citenamefont {Gunter}, \citenamefont {Skinner}, \citenamefont {Ceder},\ and\
  \citenamefont {Persson}}]{Jain2013}%
  \BibitemOpen
  \bibfield  {author} {\bibinfo {author} {\bibfnamefont {A.}~\bibnamefont
  {Jain}}, \bibinfo {author} {\bibfnamefont {S.~P.}\ \bibnamefont {Ong}},
  \bibinfo {author} {\bibfnamefont {G.}~\bibnamefont {Hautier}}, \bibinfo
  {author} {\bibfnamefont {W.}~\bibnamefont {Chen}}, \bibinfo {author}
  {\bibfnamefont {W.~D.}\ \bibnamefont {Richards}}, \bibinfo {author}
  {\bibfnamefont {S.}~\bibnamefont {Dacek}}, \bibinfo {author} {\bibfnamefont
  {S.}~\bibnamefont {Cholia}}, \bibinfo {author} {\bibfnamefont
  {D.}~\bibnamefont {Gunter}}, \bibinfo {author} {\bibfnamefont
  {D.}~\bibnamefont {Skinner}}, \bibinfo {author} {\bibfnamefont
  {G.}~\bibnamefont {Ceder}}, \ and\ \bibinfo {author} {\bibfnamefont {K.~A.}\
  \bibnamefont {Persson}},\ }\href@noop {} {\bibfield  {journal} {\bibinfo
  {journal} {APL Mater.}\ }\textbf {\bibinfo {volume} {1}},\ \bibinfo {pages}
  {011002} (\bibinfo {year} {2013})}\BibitemShut {NoStop}%
\bibitem [{\citenamefont {De~Jong}\ \emph {et~al.}(2015)\citenamefont
  {De~Jong}, \citenamefont {Chen}, \citenamefont {Angsten}, \citenamefont
  {Jain}, \citenamefont {Notestine}, \citenamefont {Gamst}, \citenamefont
  {Sluiter}, \citenamefont {Ande}, \citenamefont {Van Der~Zwaag}, \citenamefont
  {Plata} \emph {et~al.}}]{de2015charting}%
  \BibitemOpen
  \bibfield  {author} {\bibinfo {author} {\bibfnamefont {M.}~\bibnamefont
  {De~Jong}}, \bibinfo {author} {\bibfnamefont {W.}~\bibnamefont {Chen}},
  \bibinfo {author} {\bibfnamefont {T.}~\bibnamefont {Angsten}}, \bibinfo
  {author} {\bibfnamefont {A.}~\bibnamefont {Jain}}, \bibinfo {author}
  {\bibfnamefont {R.}~\bibnamefont {Notestine}}, \bibinfo {author}
  {\bibfnamefont {A.}~\bibnamefont {Gamst}}, \bibinfo {author} {\bibfnamefont
  {M.}~\bibnamefont {Sluiter}}, \bibinfo {author} {\bibfnamefont {C.~K.}\
  \bibnamefont {Ande}}, \bibinfo {author} {\bibfnamefont {S.}~\bibnamefont {Van
  Der~Zwaag}}, \bibinfo {author} {\bibfnamefont {J.~J.}\ \bibnamefont {Plata}},
   \emph {et~al.},\ }\href@noop {} {\bibfield  {journal} {\bibinfo  {journal}
  {Sci. Data}\ }\textbf {\bibinfo {volume} {2}},\ \bibinfo {pages} {150009}
  (\bibinfo {year} {2015})}\BibitemShut {NoStop}%
\end{thebibliography}%

\clearpage
\widetext
\begin{center}
\textbf{\large Supplemental Material: Stability of Electrodeposition at Solid-Solid Interfaces and Implications for Metal Anodes}
\end{center}
\setcounter{equation}{0}
\setcounter{figure}{0}
\setcounter{table}{0}
\setcounter{page}{1}
\makeatletter
\renewcommand{\theequation}{S\arabic{equation}}
\renewcommand{\thefigure}{S\arabic{figure}}

\section{Stability Criteria}
The stability parameter $\chi$ is obtained from the equation $\Delta \mu_{e^-} = \chi\text{Re}\{Ae^{ikx}\}$. Since $\Delta \mu_{e^-}$ consists of surface tension, hydrostatic and deviatoric stress terms, $\chi$ can be broken down into three terms:
\begin{gather*}\label{eq:chi}
\begin{split}
\chi=&\underbrace{-\frac{\gamma  k^2 (V_\mathrm{{M}} + V_\mathrm{{M^{z+}}})}{2 \mathrm{z}}}_\text{surface tension} \\
&+ \underbrace{\frac{ 2G_e G_s k(V_\mathrm{{M}} + V_\mathrm{{M^{z+}}}) (\nu_e (4 \nu_s-3)-3 \nu_s+2)}{\mathrm{z} (G_e (\nu_e-1) (4 \nu_s-3)+G_s (4 \nu_e-3) (\nu_s-1))}}_\text{deviatoric stress}\\
&+\underbrace{\frac{k  (V_\mathrm{{M}} - V_\mathrm{{M^{z+}}}) \left(G_e^2 (4 \nu_s-3)+G_s^2 (3-4 \nu_e)\right)}{2 \mathrm{z} (G_e (\nu_e-1) (4 \nu_s-3)+G_s (4 \nu_e-3) (\nu_s-1))}}_\text{hydrostatic stress}
\end{split}
\end{gather*}
The critical shear modulus ratio $G_s/G_e$ above (for $v>1$) or below which (for $v<1$) the electrodeposition is stable can be calculated by setting the stability parameter $\chi$ to zero, and is given by:
\begin{eqnarray*}\label{eq:gcrit}
\frac{G_s}{G_e}= 
\begin{cases}
    \frac{B+\sqrt{D}}{(v-1)(4\nu_e-3)},& \text{if } v\leq 1\\
    \frac{B-\sqrt{D}}{(v-1)(4\nu_e-3)},              & \text{if } v> 1
\end{cases}
\end{eqnarray*}
where:
\begin{eqnarray*}
B=-4 - 4 v + 6 \nu_e + 6 v \nu_e + 6 \nu_s +  6 v \nu_s - 8 \nu_e \nu_s - 8 v \nu_e \nu_s\\
 D=(v-1)^2 (-3 + 4 \nu_e) (-3 + 4 \nu_s) +  4 (v+1)^2 (2 - 3 \nu_s + \nu_e (-3 + 4 \nu_s))^2.
\end{eqnarray*}
As mentioned in the text, the surface tension term was ignored due to to its small contribution at wave number of interest.

\section{Calculation of Partial Molar Volume Ratio}
The intrinsic volume of the ion in a binary solid electrolyte $\mathrm{M_pX_q}$ can be said to follow the additivity of volumes \cite{marcus-ionvolumes}:
\begin{eqnarray*}
V_{\mathrm{total}}=pV_{\mathrm{M^{q+}}}+qV_{\mathrm{X^{p-}}}.
\end{eqnarray*} 
Further, we assumed the ratio of volumes occupied by each ion to follow:
\begin{eqnarray*}
\frac{ V_{\mathrm{M^{q+}}} }{ V_{\mathrm{X^{p-}}} } = \frac{r^3_{\mathrm{M^{q+}}}}{r^3_{\mathrm{X^{p-}}}}.
\end{eqnarray*}
where $r$ is the ionic radius of the respective ion tabulated by Shannon \cite{shannon1976revised} for monoatomic ions and by Marcus et al. \cite{marcus-ionvolumes} for multiatomic species like $\mathrm{PO_4^{3-}}$. This equation can be extended to alloys with ionic radius replaced by the atomic radius.


\end{document}